\documentclass{INTERSPEECH2023}

\interspeechcameraready 

\usepackage{bm}
\usepackage{graphicx}
\usepackage{float}
\usepackage{ascmac} 
\usepackage{fancybox}
\usepackage{float}
\usepackage{tabularx}
\usepackage{booktabs}
\usepackage{amssymb}
\usepackage{multirow}
\usepackage{cite}
\usepackage{multirow}
\usepackage[table,xcdraw]{xcolor}
\usepackage[normalem]{ulem}
\useunder{\uline}{\ul}{}
\usepackage{url}

\def\x{{\bm x}}
\def\paramssl{{\bm \theta_{\text{ssl}}}}
\def\SSL{{\tt SSL}}
\def\WS{{\tt WS}}
\def\F{{\bm F}}
\def\inRT{\in \mathbb{R}^{T}}
\def\inRdT{\in \mathbb{R}^{D \times T'}}
\def\Fn{\F_n}
\def\Fw{\F^{\tau}}
\def\wn{w^{\tau}_n}
\def\w{{\bm \theta}_{\text{w}}^{\tau}}
\def\l{\bm l}
\def\lk{\l^{\tau}}
\def\lhatk{\hat{\l}^{\tau}}
\def\DOWNSTREAMk{{\tt DOWNSTREAM}^{\tau}}

\def\paramdsk{{\bm \theta_{\text{ds}}^{\tau}}}
\def\SE{{\tt SE}}
\def\y{{\bm y}}
\def\shat{\hat{{\bm x}}}
\def\s{{\bm x}}
\def\paramse{{\bm \theta_{\text{se}}}}
\def\L{{\cal L}}
\def\Lsnr{\L_{\text{SNR}}}
\def\LSSL-MSE{\L_{\text{SSL-MSE}}}

\AtBeginDocument{
  \abovedisplayskip     =.5\abovedisplayskip
  \abovedisplayshortskip=.5\abovedisplayshortskip
  \belowdisplayskip     =.5\belowdisplayskip
  \belowdisplayshortskip=.5\belowdisplayshortskip}

\title{Downstream Task Agnostic Speech Enhancement with \\Self-Supervised Representation Loss}
\name{Hiroshi Sato, Ryo Masumura, Tsubasa Ochiai, Marc Delcroix, \\ Takafumi Moriya, Takanori Ashihara, Kentaro Shinayama, \\ Saki Mizuno, Mana Ihori, Tomohiro Tanaka, Nobukatsu Hojo}
\address{
  NTT Corporation, Japan}
\email{hrs.sato@ntt.com}

\begin{document}

\maketitle
 
\begin{abstract}
Self-supervised learning (SSL) is the latest breakthrough in speech processing,
especially for label-scarce downstream tasks by leveraging massive unlabeled audio data.
The noise robustness of the SSL is one of the important challenges to expanding its application. 
We can use speech enhancement (SE) to tackle this issue. 
However, the mismatch between the SE model and SSL models potentially limits its effect.
In this work, we propose a new SE training criterion that minimizes the distance between clean and enhanced signals in the feature representation of the SSL model to alleviate the mismatch.
We expect that the loss in the SSL domain could guide SE training to preserve or enhance various levels of characteristics of the speech signals that may be required for high-level downstream tasks.
Experiments show that our proposal improves the performance of an SE and SSL pipeline on five downstream tasks with noisy input while maintaining the SE performance.
\end{abstract}

\section{Introduction}
In the past decade, advances in deep learning have drastically improved speech processing technology and its real-world implementation.
The supervised learning framework has been a central component of these advancements, as it improves the performance in proportion to the amount of available labeled data~\cite{amodei2016deep,hinton2012deep}.
However, the annotation cost of labeled data constrains performance, particularly in tasks where a large amount of labeled data is either unavailable or too expensive to obtain.

To address this issue, self-supervised learning (SSL), a machine learning technique that can learn useful representations without the need for artificially labeled data, is attracting research attention~\cite{mohamed2022self,liu2021self,jaiswal2020survey}.
One research goal of SSL in speech processing is to obtain a single SSL model whose task-agnostic representations are effective for various downstream tasks~\cite{chung2016audio,chung2018speech2vec,schneider2019wav2vec,baevski2019vq,chung2021w2v}.
To this end, various techniques have been proposed, among which wav2vec 2.0~\cite{baevski2020wav2vec}, HuBERT~\cite{hsu2021hubert}, and WavLM~\cite{chen2022wavlm} are some of the most widely applied approaches. Indeed they provide effective representations for various downstream tasks~\cite{yang2021superb,tsai2022superb}. 

The noise robustness of the SSL is one of the critical challenges to expanding its application, although this issue has not yet been fully examined.
Among the leading SSL models, WavLM~\cite{chen2022wavlm} addresses the issue of robustness against noise and overlapping speech from interfering background speakers by incorporating a denoising framework into the pre-training process.
The prior study showed that it was relatively robust to noise ~\cite{masuyama2022end}.
Besides, Chung {\it et al} revealed that the noise robustness of SSL models could be further improved by introducing a single-channel speech enhancement (SE) frontend~\cite{chang2022end}.
They showed that the SE, SSL, and automatic speech recognition (ASR) pipeline achieves state-of-the-art performance on the monaural CHiME-4 task.
However, they also reported that the fine-tuning of SE frontend on the downstream ASR task is critically necessary to benefit from the SE frontend.
In other words, the SE model trained separately from SSL models with the SE training objective suffers from a mismatch with SSL models, potentially limiting SE performance.

To address this issue, we propose an SSL Mean Square Error (SSL-MSE) loss as way of learning an SE model optimal as the frontend of the SSL model while being independent of the downstream tasks, e.g., ASR.
Contrary to the conventional SE training objective that minimizes the distance between enhanced speech and clean source in the time or spectrum domain, SSL-MSE aims to bring these signals closer in the feature domain projected by the SSL model, which makes enhanced signal well suited for the SSL model.
Moreover, we propose using a multi-task training objective that combines the SSL-MSE loss with a conventional SE loss to preserve SE capabilities of the front-end.
Since SSL models have been shown to learn effective representations for many downstream tasks, they can capture not only acoustic but also higher-level information such as phonetic or semantic information~\cite{pasad2021layer,dunbar21_interspeech,hsieh21_interspeech}. 
Thus, we expect that creating a loss term in the SSL domain could guide SE training to preserve or enhance various levels of characteristics of the speech signals that may be required for high-level downstream tasks.

A major difference between SSL-MSE training and the fine-tuning approach of the SE model with a downstream task adopted in~\cite{chang2022end} is that ours can train the SE model in a downstream task agnostic manner.
In \cite{chang2022end}, the SE frontend and downstream ASR models are jointly trained on the ASR criterion while freezing the SSL model.
On the other hand, our proposal constructs a single SE model suitable for a general SSL pipeline that can be applied to various downstream tasks without requiring downstream task-specific fine-tuning.
This allows sharing a single SE and SSL model for multiple downstream tasks by only training the relatively small-sized downstream model for each task.

We confirmed the effectiveness of the proposed SSL-MSE loss with various downstream tasks by performing experiments on a noisy version of a subset of the SUPERB benchmark~\cite{yang2021superb}
The result shows that the proposed SSL-MSE substantially improves the performance of downstream tasks using WavLM.
Our proposal is also effective even with noise-robust downstream models trained with noisy paired data.
Moreover, when using the multitask loss, the proposed SE system maintains SE performance, allowing using it also for applications requiring access to the enhanced speech signal, e.g., listening applications.
It can also contribute to the interpretability of the process.
An interesting finding of our study is that SE systems trained with SSL-MSE loss generalize to downstream tasks using different SSL models from those used for the SSL-MSE training. This seems to confirm our intuition that the task-agnostic feature representation of SSL models helps us to learn a more robust SE frontend.


\section{Conventional Method}
\subsection{Task-agnostic SSL upstream models}
Various types of SSL methods have been proposed to extract powerful features or representations from speech recordings that are useful for downstream tasks.
We can write the feature extraction process with the SSL model as follows:
\begin{align}
    \label{eq:ssl}
    \F_{1:N} = \SSL(\x;\paramssl),
\end{align}
where $\x \inRT$ denotes the monaural input signal in raw waveform with $T$ samples, $\F_n \inRdT$ denotes the time series of extracted features obtained from the $n$-th layer of the SSL model, $N$ is the number of layers of the SSL model, and $\paramssl$ denotes the learnable parameters of the SSL model. $T'$ and $D$ denote the number of frames and dimensions of the extracted features, respectively.
The learnable parameters, $\paramssl$ are trained using pretext tasks with unpaired audio-only data.
For example, HuBERT~\cite{hsu2021hubert} and WavLM~\cite{chen2022wavlm} are trained using the BERT-like masked prediction task~\cite{devlin2018bert} on the target label generated in an offline clustering step.
Of particular note, WavLM is made robust to noise and interference speakers through the addition of Deep Noise Suppression (DNS) noise~\cite{reddy2021interspeech} and interference speakers during its training.
Various types of pre-trained SSL models have been made publicly available for application to downstream tasks.

\subsection{Task specific downstream models}
There are two major ways to apply SSL models to downstream tasks, either as a fixed feature extractor, or permitting retraining of their parameters~\cite{mohamed2022self}.
Following previous studies~\cite{yang2021superb,chang2022end,masuyama2022end}, we adopted the former approach, i.e., freezing the parameters of the SSL model $\paramssl$ during the training of the downstream models, because SSL models are usually very large and thus, it is computationally too intensive to fine-tune the SSL models for every downstream task.
The downstream task-specific additional layers are relatively small deep neural networks, which we call the downstream model.

An effective way to apply the SSL features across various types of downstream tasks is to use the weighted sum of the embeddings from different layers in SSL model as the input feature of the downstream model~\cite{chen2022wavlm, yang2021superb}.
The process of the downstream model for a task $\tau$ can be written as follows:
\vspace{-2pt}
\begin{align}
    \label{eq:weighting}
    \Fw & = \WS(\F_{1:N};\w)
     \equiv \sum_{n=1}^{N} \wn\Fn, \\
    \label{eq:downstream}
    \lhatk & = \DOWNSTREAMk(\Fw;\paramdsk),
\end{align}
where $\WS(\cdot)$ represents the weighted sum function, $\w = [w_1^{\tau},...,w_N^{\tau}]$ are the weights for latent representations obtained from each layer, which are learnable parameters, and $\lhatk$ is the estimation result attained by the downstream model. 
$\DOWNSTREAMk(\cdot)$ denotes the downstream model for the task $\tau$ whose learnable parameters are $\paramdsk$.
The learnable parameters $(\w,\paramdsk)$ are jointly optimized by using task-specific paired data $(\lk, \lhatk)$, while the upstream model $\paramssl$ is frozen.

\subsection{Speech Enhancement and SSL pipeline}
Speech enhancement is a method to suppress undesired audio from the input noisy observations.
We denote the SE process as follows:
\vspace{-11pt}
\begin{align}
    \label{eq:se}
    \shat = \SE(\y; \paramse),
\end{align}
where $\y \inRT$ and $\shat \inRT$ denote the noisy observations and enhanced signals, respectively. 
$\paramse$ denotes the learnable parameters of the SE model, which is optimized by minimizing the distance between enhanced signal $\shat$ and the ground truth clean source $\s \inRT$.
Specifically, we adopt the scale-dependent signal-to-noise ratio loss $\Lsnr(\cdot)$ as the distance measure, which is defined as follows:
\vspace{-3pt}
\begin{align}
    \label{eq:sdsnr}
    \Lsnr = 10\log_{10}\frac{\lVert\shat\rVert^2}{\lVert\s-\shat\rVert^2}.
\end{align}
Since the clean source is not available for real-recorded noisy speech, simulated mixtures of clean source $\s$ and noise are commonly used as noisy speech $\y$ in SE training.
The entire SE and SSL pipeline for downstream task $\tau$ can be written as follows:
\begin{align}
    \label{eq:pipeline}
     \lhatk = \DOWNSTREAMk(\WS(\SSL(\SE(\y; \paramse);\paramssl);\w);\paramdsk).
\end{align}

It is commonly known that the mismatch between SE frontend and backends can degrade the performance of backend tasks when the SE frontend is trained separately from backends, using the SE training criterion~\cite{iwamoto2022bad,sato2022learning,sato2021should}.
For the SE, SSL, and ASR pipeline, this mismatch was mitigated by using the downstream ASR loss in jointly retraining SE and downstream models~\cite{chang2022end, masuyama2022end}.
However, this method may not be feasible when considering application for more than a few downstream tasks, because the retraining of the SE model for every downstream task is computationally demanding as the SE model tend to have many more learnable parameters than those of the downstream models to maintain the enhancement performance.
Moreover, the fine-tuning on the downstream task makes the SE frontend task-dependent, which makes it infeasible to share a single general SE model for multiple downstream tasks.


\begin{figure}[tb]
 \begin{center}
  \includegraphics[width=.85\hsize]{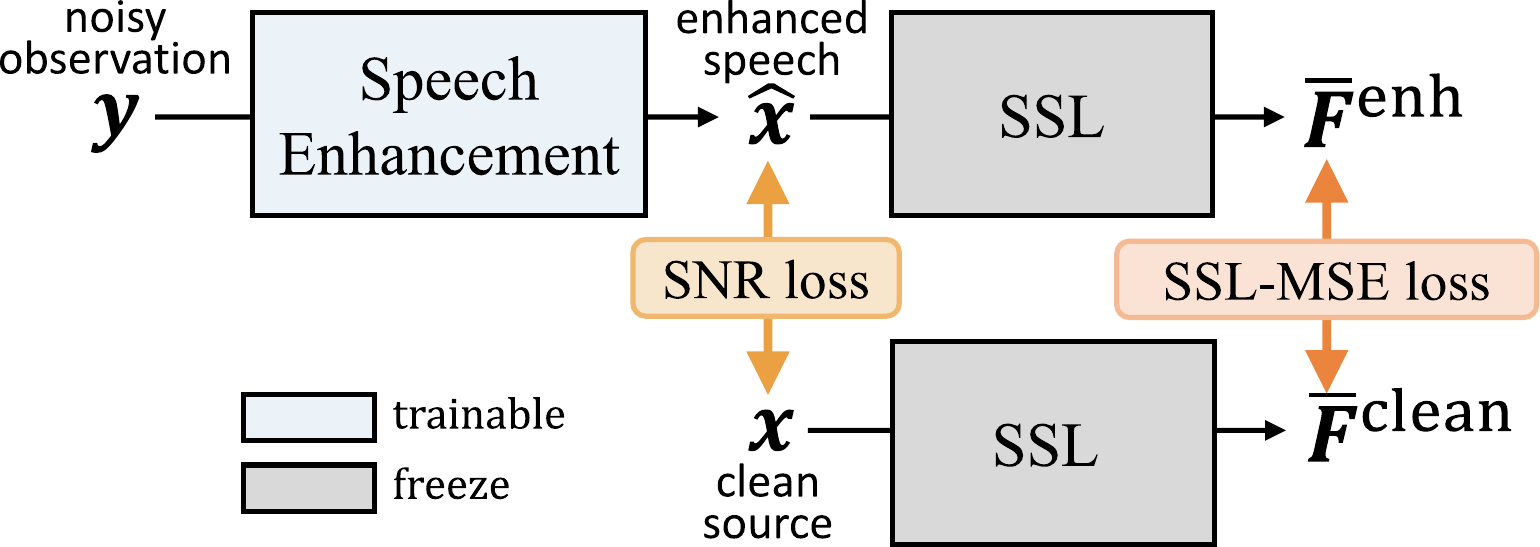}
 \end{center}
 \vspace{-12pt}
 \caption{Overview of the proposed SSL-MSE loss.}
 \vspace{-16pt}
 \label{fig:framework}
\end{figure}

\section{Proposed Method}
\label{sec:proposed}
To obtain the downstream task-agnostic SE frontend while reducing the mismatch, we propose SSL-MSE.
Figure~\ref{fig:framework} shows the general framework of the proposed SSL-MSE training.
In addition to the conventional signal-to-noise ratio (SNR) loss, we calculate SSL-MSE which drives the enhanced signal closer to the clean source in the SSL feature domain.
Formally, SSL-MSE $\LSSL-MSE$ between enhanced speech $\shat=\SE(\y; \paramse)$ and clean source $\s$ is calculated as the mean square error between SSL features extracted from these signals as follows:
\begin{align}
    \label{eq:SSL-MSE}
    \LSSL-MSE = \lVert\overline{{\bm F}}^{\text{enh}} - \overline{{\bm F}}^{\text{clean}}\rVert_{\hspace{-1pt}F}^2/DT',\hspace{12mm} \\
    \overline{{\bm F}}^{\text{enh}} = \sum_{n=1}^{N}\Tilde{w}_n\F_n^{\text{enh}}, \hspace{3mm}
    \overline{{\bm F}}^{\text{clean}} \hspace{-1mm}= \sum_{n=1}^{N}\Tilde{w}_n\F_n^{\text{clean}}, \hspace{5mm} \\
    \F_{1:N}^{\text{enh}} = \SSL(\SE(\y; \paramse);\paramssl),
    \hspace{2mm}
    \F_{1:N}^{\text{clean}} = \SSL(\s;\paramssl),
\end{align}
where $\lVert\cdot\rVert_{\hspace{-1pt}F}$ represents Frobenius norm and $\tilde{\bm{w}} = [\tilde{w}_1,...,\tilde{w}_N]$ represents the layer weight for the latent feature obtained from each layer of the SSL model. 
The layer weight is a hyperparameter that controls which layer output SSL-MSE focuses on.
We tested three types of weighting: 1) {\it last} uses only the last layer output, 2) {\it all} weighs every layer equally, and 3) {\it latter-half} weights uniformly over the latter half layers.
The model parameter $\paramse$ is optimized by minimizing the multitask loss function $\L$ expressed as follow:
\begin{align}
    \label{eq:overallloss}
    \L = \LSSL-MSE + \alpha\Lsnr,
\end{align}
where $\alpha$ denotes the multitask weight for the SNR loss~\cite{vincent2006performance}. We hereafter refer to the hyperparameter $\alpha$ as SNR weight.
Note that the parameters of the SSL model $\paramssl$ are frozen while the SSL-MSE training.

We adopted WavLM~\cite{chen2022wavlm} as a SSL model to calculate SSL-MSE loss.
Note that SSL-MSE training does not depend on the downstream tasks, only on the SSL model. 
How the SSL-MSE training generalizes over other types of SSL models than that used for SSL-MSE training is discussed in Section~\ref{sec:result}.


\section{Experiments}
\subsection{Experimental setup}
\subsubsection{Speech enhancement frontend}
The SE model was trained on simulated mixtures of speech and noise data.
We used LibriSpeech~\cite{panayotov2015librispeech} for speech recordings and DNS noise~\cite{reddy2021interspeech} for noise recordings.
The number of noisy observations were 100,000 and 5,000 for the training and development sets, respectively.
The noise was added at SNR values randomly sampled from -3 to 20 dB.
We chose DNS noise to create a fair comparison of the SSL pipeline with/without the SE module, as DNS noise is also used in training the WavLM models.
We adopted ConvTasnet as the SE front-end module, which converts noisy raw-waveform audio into enhanced raw-waveforms in a time-domain, end-to-end processing manner~\cite{luo2019conv}. 
According to the setup adopted in~\cite{e3net}, we set the hyperparameters to ${\rm N}\hspace{-1mm}=\hspace{-1mm}4096$, ${\rm L}\hspace{-1mm}=\hspace{-1mm}320$, ${\rm B}\hspace{-1mm}=\hspace{-1mm}256$, ${\rm R}\hspace{-1mm}=\hspace{-1mm}4$, ${\rm X}\hspace{-1mm}=\hspace{-1mm}8$, ${\rm H}\hspace{-1mm}=\hspace{-1mm}512$ and ${\rm P}\hspace{-1mm}=\hspace{-1mm}3$ following the notation in \cite{luo2019conv}.
The SSL-MSE loss was introduced after pre-training the model with conventional SNR loss to speed up the conversion.
The initial learning rate for the pre-training with SNR loss was set to 5e-4 and was multiplied by 3/4 if the loss on the development set did not decrease for 2 epochs. 
For optimization, we adopted the Adam optimizer\cite{kingma2014adam}. 
The models were pre-trained for 100 epochs.

\subsubsection{SSL-MSE loss}
To calculate SSL-MSE, we used off-the-shelf WavLM {\sc Base+} model.
We tested three configuration of layer weight $\tilde{\bm{w}}$: last, all, and latter-half as defined in Section~\ref{sec:proposed}. We used the last setup unless otherwise specified.
We tested the SNR weight $\alpha$ within $\{0,0.0001,0.001,0.01,0.1,1\}$. 
We adopted 0.1 unless otherwise specified, which is the smallest value where SNR loss on development set did not degrade compared with that of conventional SNR loss training.
We fine-tuned the SE model using SSL-MSE loss with the initial learning rate of 1e-4 and for up to 50 epochs.

\subsubsection{SSL and downstream model}
We implemented the SSL pipeline using the S3PRL toolkit~\cite{yang2021superb,s3prl}. 
As the SSL model, we used the publicly available pre-trained models of wav2vec 2.0 {\sc Base}, Hubert {\sc Base}, WavLM {\sc Base+}, and WavLM {\sc Large} without updating their parameters.
We tested two setups for downstream model training: 1) {\it official} setup where downstream models were trained with SUPERB official training data that is relatively `clean' speech data, according to S3PRL SUPERB recipe~\cite{s3prl}, and 2) {\it noise robust} setup where DNS noise was added to SUPERB official downstream training data at SNR values randomly sampled from -3 to 20 dB. 
The results gained in the noise robust setup are shown in Table~\ref{tab:noiserobust}; for the other experiments we adopt the official setup.
SE frontend was not applied while training the downstream model in noise robust setup as well as in official setup.
The downstream models were prepared for each downstream task for each SSL model.

\subsubsection{Evaluation details}
In this work, we evaluate the SE model in terms of 1) speech enhancement performance and 2) the performance of downstream tasks.
SE performance was measured using the 3,000 simulated noisy observations of Librispeech data and DNS noise at SNR values randomly sampled from 0 to 10 dB. 
As the performance measure, we adopted average value of the scale-independent source-to-distortion ratio (SDR)~\cite{le2019sdr} and perceptual evaluation of speech quality (PESQ), as well as SSL-MSE values calculated on the features obtained from the last layer of the SSL model.
The performance of downstream tasks was evaluated on a subset of the SUPERB benchmark.
Namely, we chose ASR, phone recognition (PR), automatic speaker verification (ASV), intent classification (IC), and slot-filling (SF) tasks to cover three major categories of tasks, i.e., content, speaker, and semantics. ASR performance was measured on the test-clean set and without the language model.
To evaluate the noise robustness, we prepared a noisy version of the SUPERB test sets for each task by adding DNS noise to the original recordings at SNR values randomly sampled from 0 to 10 dB.
Since DNS noise contains a wide variety of the noise types and we use different samples for training and testing, we can verify robustness to unseen noise although we adopted DNS noise in both training and evaluation.
\begin{table*}[tb]
\centering
\caption{The performance of each system evaluated by SE performance and downstream performance with \textbf{\textit{WavLM Base+}}. For the downstream performance, row (a1) shows the result for official SUPERB eval sets and (a2)-(a7) for noisy eval sets. (a2) is the result of noisy speech without front-end processing. (a3) indicates performance with conventional SE model and (a4)-(a7) indicate that with SE trained with proposed SSL-MSE. Column `SSL-MSE' shows the distance between enhanced or noisy speech and clean source measured on the SSL feature obtained from the last layer. SSL-MSE and the SE performance are evaluated on simulated noisy observation of Librispeech and DNS noise.}
\vspace{-7pt}
\label{tab:main}
\scalebox{0.88}[0.88]{
\begin{tabular}{cccccc||ccccc|c|cc}
\hline
 & \multirow{3}{*}{\begin{tabular}[c]{@{}c@{}}Test\\ set\end{tabular}} & \multirow{3}{*}{\begin{tabular}[c]{@{}c@{}}SE\\ frontend\end{tabular}} & \multirow{3}{*}{\begin{tabular}[c]{@{}c@{}}SSL\\ MSE\\ Loss\end{tabular}} & \multirow{3}{*}{\begin{tabular}[c]{@{}c@{}}Layer\\ weight\\ $\tilde{\bm{w}}$\end{tabular}} & \multirow{3}{*}{\begin{tabular}[c]{@{}c@{}}SNR\\ weight\\ $\alpha$\end{tabular}} & \multicolumn{5}{c|}{Downstream} & \multirow{2}{*}{\begin{tabular}[c]{@{}c@{}}SSL\\ MSE\end{tabular}} & \multicolumn{2}{c}{\multirow{2}{*}{\begin{tabular}[c]{@{}c@{}}Speech\\ Enhancement\end{tabular}}} \\ \cline{7-11}
 &  &  &  &  &  & \multicolumn{1}{c|}{ASR} & \multicolumn{1}{c|}{PR} & \multicolumn{1}{c|}{ASV} & \multicolumn{1}{c|}{IC} & SF &  & \multicolumn{2}{c}{} \\ \cline{7-14} 
 &  &  &  &  &  & \multicolumn{1}{c|}{WER↓} & \multicolumn{1}{c|}{PER↓} & \multicolumn{1}{c|}{EER↓} & \multicolumn{1}{c|}{Acc↑} & F1↑ & MSE↓ & \multicolumn{1}{c|}{SDR↑} & PESQ↑ \\ \hline
(a1) & clean &  &  &  &  & \multicolumn{1}{c|}{5.6} & \multicolumn{1}{c|}{4.4} & \multicolumn{1}{c|}{4.5} & \multicolumn{1}{c|}{98.8} & 90.3 & - & \multicolumn{1}{c|}{-} & - \\ \hline
(a2) & noisy &  &  &  &  & \multicolumn{1}{c|}{17.1} & \multicolumn{1}{c|}{16.0} & \multicolumn{1}{c|}{10.9} & \multicolumn{1}{c|}{67.4} & 82.4 & 0.0199 & \multicolumn{1}{c|}{3.7} & 1.27 \\ \hline
(a3) & noisy & \checkmark & No &  &  & \multicolumn{1}{c|}{14.6} & \multicolumn{1}{c|}{9.9} & \multicolumn{1}{c|}{8.6} & \multicolumn{1}{c|}{84.8} & 82.4 & 0.0122 & \multicolumn{1}{c|}{15.7} & 2.27 \\ \hline
(a4) & noisy & \checkmark & Yes & last & 0.1 & \multicolumn{1}{c|}{13.1} & \multicolumn{1}{c|}{8.8} & \multicolumn{1}{c|}{8.0} & \multicolumn{1}{c|}{87.6} & 83.5 & 0.0105 & \multicolumn{1}{c|}{\textbf{15.8}} & 2.34 \\
(a5) & noisy & \checkmark & Yes & all & 0.1 & \multicolumn{1}{c|}{13.5} & \multicolumn{1}{c|}{9.0} & \multicolumn{1}{c|}{8.0} & \multicolumn{1}{c|}{87.3} & 83.1 & 0.0106 & \multicolumn{1}{c|}{\textbf{15.8}} & \textbf{2.35} \\
(a6) & noisy & \checkmark & Yes & latter-half & 0.1 & \multicolumn{1}{c|}{12.9} & \multicolumn{1}{c|}{8.7} & \multicolumn{1}{c|}{\textbf{7.9}} & \multicolumn{1}{c|}{\textbf{87.8}} & 83.4 & 0.0103 & \multicolumn{1}{c|}{\textbf{15.8}} & \textbf{2.35} \\
(a7) & noisy & \checkmark & Yes & last & 0 & \multicolumn{1}{c|}{\textbf{11.2}} & \multicolumn{1}{c|}{\textbf{7.3}} & \multicolumn{1}{c|}{8.4} & \multicolumn{1}{c|}{87.7} & \textbf{84.9} & \textbf{0.0093} & \multicolumn{1}{c|}{8.0} & 1.38 \\ \hline
\end{tabular}
}
\vspace{-8pt}
\end{table*}
\begin{table}[tb]
\centering
\caption{The performance with SSL models other than that used for SSL-MSE training. `-' in SE column means the performance without SE, `baseline' is a condition with the SE trained on SNR loss, and `+SSL-MSE' is the SE trained with SSL-MSE calculated on the last output of WavLM Base+ at SNR weight of 0.1.}
\vspace{-4pt}
\label{tab:mismatch}
\scalebox{0.85}[0.85]{
\begin{tabular}{ccc||c|ccc}
\hline
\multicolumn{1}{c}{} & \multirow{3}{*}{SSL} & \multirow{3}{*}{SE} & \multirow{2}{*}{\begin{tabular}[c]{@{}c@{}}SSL\\ MSE\end{tabular}} & \multicolumn{3}{c}{Downstream} \\ \cline{5-7} 
\multicolumn{1}{c}{} &  &  &  & \multicolumn{1}{c|}{ASR} & \multicolumn{1}{c|}{ASV} & IC \\ \cline{4-7} 
\multicolumn{1}{c}{} &  &  & MSE↓ & \multicolumn{1}{c|}{WER↓} & \multicolumn{1}{c|}{EER↓} & Acc↑ \\ \hline
\hspace{-1mm}(b2)\hspace{-2mm} & w2v 2.0 & - & 0.1217 & \multicolumn{1}{c|}{36.8} & \multicolumn{1}{c|}{19.0} & 41.0 \\
\hspace{-1mm}(b3)\hspace{-2mm} & w2v 2.0 & baseline & 0.0450 & \multicolumn{1}{c|}{18.8} & \multicolumn{1}{c|}{12.2} & 70.9 \\
\hspace{-1mm}(b4)\hspace{-2mm} & w2v 2.0 & \hspace{-3mm}+SSLMSE\hspace{-1mm} & \textbf{0.0428} & \multicolumn{1}{c|}{\textbf{17.5}} & \multicolumn{1}{c|}{\textbf{11.3}} & \textbf{71.3} \\ \hline
\hspace{-1mm}(c2)\hspace{-2mm} & HuBERT & - & 0.1207 & \multicolumn{1}{c|}{31.7} & \multicolumn{1}{c|}{15.5} & 57.8 \\
\hspace{-1mm}(c3)\hspace{-2mm} & HuBERT & baseline & 0.0696 & \multicolumn{1}{c|}{18.5} & \multicolumn{1}{c|}{10.6} & 81.9 \\
\hspace{-1mm}(c4)\hspace{-2mm} & HuBERT & \hspace{-3mm}+SSLMSE\hspace{-1mm} & \textbf{0.0628} & \multicolumn{1}{c|}{\textbf{16.9}} & \multicolumn{1}{c|}{\textbf{9.7}} & \textbf{82.6} \\ \hline
\hspace{-1mm}(d2)\hspace{-2mm} & WavLM L & - & 0.0281 & \multicolumn{1}{c|}{9.9} & \multicolumn{1}{c|}{9.8} & 52.1 \\
\hspace{-1mm}(d3)\hspace{-2mm} & WavLM L & baseline & 0.0154 & \multicolumn{1}{c|}{7.8} & \multicolumn{1}{c|}{9.8} & 86.6 \\
\hspace{-1mm}(d4)\hspace{-2mm} & WavLM L & \hspace{-3mm}+SSLMSE\hspace{-1mm} & \textbf{0.0138} & \multicolumn{1}{c|}{\textbf{7.2}} & \multicolumn{1}{c|}{\textbf{9.3}} & \textbf{88.3} \\ \hline
\end{tabular}
}
\vspace{-8pt}
\end{table}

\begin{table}[tb]
\centering
\vspace{-3pt}
\caption{The performance for noise-robust downstream models. WavLM Base+ is used to obtain the result. (a2) in this figure corresponds to system (a2) in Table~\ref{tab:main}.}
\vspace{-8pt}
\label{tab:noiserobust}
\scalebox{0.88}[0.88]{
\begin{tabular}{ccc||ccc}
\hline
 & \multirow{3}{*}{\begin{tabular}[c]{@{}c@{}}Noise\\ robust\\ downstream\end{tabular}} & \multirow{3}{*}{SE} & \multicolumn{3}{c}{Downstream} \\ \cline{4-6} 
 &  &  & \multicolumn{1}{c|}{ASR} & \multicolumn{1}{c|}{ASV} & IC \\ \cline{4-6} 
 &  &  & \multicolumn{1}{c|}{WER↓} & \multicolumn{1}{c|}{EER↓} & Acc↑ \\ \hline
(a2) &  & - & \multicolumn{1}{c|}{17.1} & \multicolumn{1}{c|}{10.9} & 67.4 \\ \hline
(e2) & \checkmark & - & \multicolumn{1}{c|}{13.6} & \multicolumn{1}{c|}{9.2} & 79.4 \\
(e3) & \checkmark & baseline & \multicolumn{1}{c|}{12.5} & \multicolumn{1}{c|}{7.9} & 85.8 \\
(e4) & \checkmark & +SSLMSE & \multicolumn{1}{c|}{\textbf{11.3}} & \multicolumn{1}{c|}{\textbf{7.7}} & \textbf{88.3} \\ \hline
\end{tabular}
}
\vspace{-12pt}
\end{table}

\subsection{Experimental results}
\subsubsection{Performance comparison}
\label{sec:result}
Table \ref{tab:main} shows the performance of the SE model and the SSL pipeline with the SE frontend.
Compared with the results gained from the clean test sets (a1), the addition of noise substantially degrades the performance of SSL pipeline as shown in (a2).
Although WavLM {\sc Base+} is trained robustly to DNS noise, the baseline SE frontend still improves the downstream performance in 4 out of 5 tasks (a3).

Column `SSL-MSE' in the table indicates that by introducing the SE frontend trained with SSL-MSE (a4)-(a7), the output features of the SSL model for noisy speech become closer to those for clean speech compared with that of the baseline SE frontend (a3). 
Accordingly, the proposed systems considerably improved the performance of the downstream tasks compared with the baseline.
The system (a4)-(a6) show that SSL-MSE training also slightly improved the SDR and PESQ compared with the baseline when the SNR weight of SSL-MSE training is set as 0.1. 
The system trained only with SSL-MSE loss (a7) shows the best performance in three out of five downstream tasks, while SDR and PESQ largely drop compared with system (a4)-(a6).
This indicate that there is a tradeoff between SE performance and the compatibility with the SSL models, which we further discuss about in Section~\ref{sec:result2}.
As for the layer weight for the SSL-MSE, the latter-half setup shows superior performance on average (a6). 
Summarizing these results, the proposed SSL-MSE can successfully train the downstream task-agnostic SE model that improves the noise robustness of the SSL pipeline while maintaining the SE performance itself.

Table \ref{tab:mismatch} shows the performance for SSL pipelines with SSL models other than WavLM {\sc Base+} which is used for SSL-MSE training.
The results show that SSL-MSE improves the performance of the SSL pipeline even for the mismatched SSL models that are unseen during training, compared with the baseline SE training.
Thus, it can be said that SSL-MSE training potentially generalizes over different types of SSL models.

Table \ref{tab:noiserobust} shows the performance of SSL pipelines when the downstream models are also trained to be robust to noise.
As seen by comparing (a2) to (e2), the noise-robust training of the downstream model is effective for the noise-robust WavLM model. 
Row (e3) shows that the SE frontend is effective even with the noise-robust SSL model and downstream models.
Furthermore, introducing the proposed SSL-MSE training (e4), achieves an additional performance improvement.
This result indicates that SSL-MSE and the noise robust training of SSL model and downstream models are complementary and can be combined to obtain superior performance.

\subsubsection{Further discussion about the effect of SSL-MSE}
\label{sec:result2}
\begin{figure}[tb]
 \begin{center}
  \includegraphics[width=0.95\hsize]{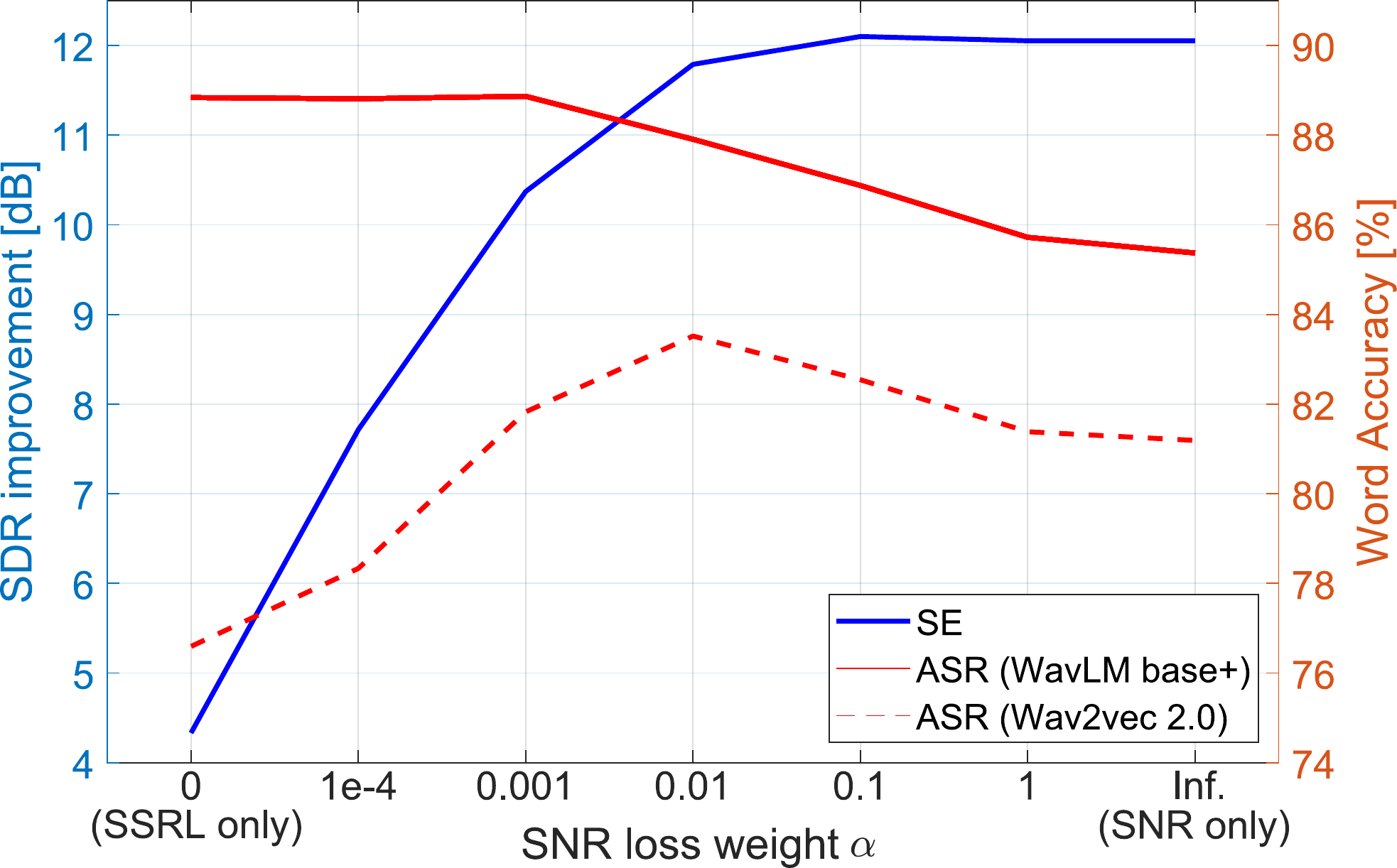}
 \end{center}
 \vspace{-16pt}
 \caption{The tradeoff between enhancement and downstream ASR performance as a function of SNR loss weight. The results for WavLM Base+ and Wav2vec 2.0 are shown.}
 \vspace{-10pt}
 \label{fig:tradeoff}
\end{figure}

To further understand the effect of the combination of SNR loss and SSL-MSE, we measured the SE and downstream ASR performance obtained with SE models trained with different SNR weights $\alpha$.
Figure \ref{fig:tradeoff} shows that the downstream ASR performance with the WavLM {\sc Base+} model, shown by the red solid line, increases as SSL-MSE becomes even more dominant than SNR loss, while the performance of SE, shown by the blue solid line, decreases when SNR loss weight is less than 0.1.
This indicates that SNR and SSL-MSE metrics requires different modes of enhancement and thus they are in a tradeoff relationship to some extent.
The red dotted line plots the downstream ASR performance with Wav2vec 2.0 model.
With a SSL model mismatched with that used for SSL-MSE calculation, the downstream performance decreases when SSL-MSE loss becomes significantly stronger than SNR loss.
It seems that the SNR objective works as a regularization term and is required to maintain the generalizability of the SE model over different types of SSL models. 


\vspace{-10pt}
\section{Conclusion}
To construct an SE model that can effectively be combined with SSL models with less mismatch, we proposed SSL-MSE, which is a minimization criterion involving enhanced signals and clean sources in the SSL feature domain.
The SE model trained with a multi-task loss including SSL-MSE is shown to improve the performance of downstream tasks in the combination with SSL models while retaining the SE performance itself.
In addition to the downstream task-agnostic nature of the proposed method, it is shown that SSL-MSE can generalize over different types of SSL models.

\bibliographystyle{IEEEtran}
\bibliography{mybib}

\end{document}